# Sublimation as an effective mechanism for flattened lobes of (486958) Arrokoth

Y. Zhao[1,2], L. Rezac[3], Y. Skorov[4], S.C. Hu[1,2], N. Samarasinha[5], J-Y. Li[5]

1. Key Laboratory of Planetary Sciences, Purple Mountain Observatory, Chinese Academy of Sciences, Nanjing 210008, China
2. CAS Center for Excellence in Comparative Planetology, Chinese Academy of Sciences, Hefei, 230026, China
3. Max-Planck-Institut für Sonnensystemforschung, Justus-von-Liebig-Weg 3, Göttingen 37077, Germany
4. Institut für Geophysik und extraterrestrische Physik, Technische Universität Braunschweig, Mendelssohnstr. 3, D-38106 Braunschweig, Germany
5. Planetary Science Institute, 1700 E Fort Lowell Road, Tucson, AZ 85719, USA

**The New Horizons spacecraft's flyby of Kuiper Belt Object (KBO) (486958) Arrokoth revealed a bilobed shape with highly flattened lobes both aligned to its equatorial plane, and a rotational axis almost aligned to the orbital plane (obliquity ~99º)[1–4]. Arrokoth belongs to the Cold Classical Kuiper Belt Object population that occupies dynamically undisturbed orbits around the Sun, and as such, is a primitive object that formed in situ. Therefore, whether its shape is primordial or evolutionary carries important implications for understanding the evolution of both KBOs and potentially their dynamically derived objects, Centaurs and Jupiter Family Comets (JFC)[5,6]. Applying our mass loss driven shape evolution model (MONET) [7], here we suggest that the current shape of Arrokoth could be of evolutionary orgin due to volatile outgassing in a timescale of about 1-100 Myr, while its spin state would not significantly affected. We further argue that such a process may be ubiquitous in the evolution of the shape of KBOs shortly after their formation. This shape changing process could also be reactivated when KBOs dynamically evolve to become Centaurs and then JFCs[5,6] and receive dramatically increased solar heating.**

The shape of Arrokoth was derived from New Horizons' flyby images as bilobed with respective dimensions of 20.6×19.9×9.4 km and 15.4×13.8×9.8 km, and uncertainties[2] of 0.5 km×0.5 km×2.0 km. The initial explanation for the formation of the flattened lobes supposes a primordial origin during the time of pebble accretion and low speed merging of multi-km scale planetesimals[1,3]. Here we perform simulations of shape evolution accounting for removal of material due to sublimation of ices[7] (Methods) to assess the feasibility and the role of this process in flattening the body and altering its spin characteristics. Our modeling builds on our previous findings[7] that sublimation driven mass loss induced shape evolution could be a general process that, under the influence of orbital and spin configurations, determines the overall shape of comets and KBO. The most relevant here are the necessary and sufficient conditions for objects to evolve a significant hemispheric asymmetry or become elongated or flattened. The key parameters are the obliquity, eccentricity and the profile of a functional dependence of mass loss rate versus solar flux[7].

Our simulation starts with a synthetic Arrokoth analogue shape, SHAPE-SYN (Fig. 1c). We derive this initial synthetic shape model by applying our model (MONET) inversely in time to the current shape of Arrokoth[1] while fixing the spin orientation[2] and orbital parameters[8] to the current values.

zhaoyuhui@pmo.ac.cn, rezac@mps.mpg.de

The inversely evolved shape model (SHAPE-REV. Fig. 1b) is generalized by joining a sphere and an oblate shape that have the same respective size parameters as the corresponding lobes but scaled to preserve the total volume. The resulting bilobed shape, SHAPE-SYN, is consistent with configuration of low speed mergers between two bodies[3] as revealed by numerical modeling (Extended Data Fig. 1, Supplementary Video. 1).

The fundamental thermophysical and structural properties of neither Arrokoth nor its precursor are constrained at this time. Nonetheless, our sublimation model implements a two layer thermophysical model[9] with $CH_4$ as the main ice component below a dust layer (model assumptions in Methods) which provides an opportunity to simulate first order effects of dust layer resistance to gas flow, tensile strength and gravity. Outgassing models with different dust layer properties, monomer size $R_a$ and its thickness (Fig. 2), lead to slight variations in the distribution of mass loss over the evolved body (Fig. 3). However, the end states of our simulations all reproduce the main characteristics of SHAPE-ORIG, in particular the pronounced flattening in the polar regions. The evolution time $T_e$ to reach the approximate volume of SHAPE-ORIG varies within a factor of 2.5 for presented cases (Fig. 3, Extended Data Fig. 2). The material removal induces an erosion of about 11 km along the z axis, and about 3 km in both x and y directions (Supplementary Video. 2). The flattening of Arrokoth's shape is a natural outcome due to a favorable combination of ~90° obliquity, small eccentricity, and mass loss rate variation with solar flux, resulting in nearly symmetric erosion between north and south hemispheres[7]. Due to the orientation of Arrokoth, the polar regions experience continuous solar illumination during polar days (with strong mass loss), while the equatorial regions are dominated by diurnal variations year around[4]. Therefore, the polar regions reach higher peak temperatures than the equator, due to the fact that the thermal timescales are longer than the rotational timescales. Thus, the poles experience more sublimation than the equatorial regions.

While spin and orbital parameters are fully constrained by the observations[2,8], the mass loss rate and solar flux relationship is a complicated modeling parameter depending on the assumed properties and conditions of dust and ice. Fig. 4 shows the role that gravity and the tensile strength of the insulating dust layer may play in shaping an object during its mass loss evolution. These additional effects modify the dependence of mass loss on solar flux, and consequently may introduce considerable north/south dichotomy or other features that are non-intuitive. Both the tensile strength and gravity effects are simulated by imposing a minimum threshold on the gas pressure (under the dust layer) for mass loss to take place. When the tensile strength is considered, a more pronounced flattening (Fig. 4b) or significant north/south hemispheric dichotomy (Fig. 4c) could appear. On the other hand, including gravity leads to a final shape that preserves the flattened shape while displays asymmetric pinnacle/hill at gravitational highs in both north and south hemispheres (Fig. 4d). These results are qualitative because many key parameters essential for more realistic modeling scenarios are unknown. Nevertheless, they demonstrate the potential effects of key mechanical properties on determining the large scale morphology on active bodies.

These results imply that Arrokoth should be rather homogeneous since there is a similar degree of flattening in the north and south hemispheres.

The proposed flattening mechanism requires the maintenance of Arrokoth's obliquity. We investigate the upper limits of spin state change of the evolving shape due to sublimation induced torques with a pure $CH_4$ composition for both SHAPE-SYN and SHAPE-REV. The change of spin axis orientation is expected to be less than 10º for both initial shapes to reach the approximate volume of Arrokoth in about 1 Myr (Extended Data Fig. 3). The stable spin state (Extended Data Fig. 3, 4) of Arrokoth during the evolution results from its symmetric shape and large obliquity (Extended Data Fig. 5). Although different volatile species or dust layer properties may vary the evolution time, the associated sublimation torques is proportional to the mass loss rate to the first order, hence always show a similar level of perturbations as discussed, independent of evolution time. Considering the limited collisions and planetary encounters[1–3,10] as well as the neglectable YORP effect[1,3] in this region of the Solar System, we conclude that the orientation and spin state of Arrokoth has not changed significantly in its evolutionary history.

Our analysis assumes $CH_4$ as the driving volatile for the evolution of Arrokoth's precursor, but other super volatiles or their mixtures would also lead to similar final shapes. Ices such as $N_2$, CO and $CH_4$ (ordered from the highest to lowest volatility) are known to have active sublimation and condensation cycles on large KBOs such as Triton[11–13] and Pluto[14,15]. Smaller bodies residing at or originating from the Kuiper belt (comets, centaurs) are generally expected to have these ices depleted on the surface due to sublimation early in their history[16], but could still exist at depth beneath the surface in the present[16–20]. Moreover, the $CH_3OH$ ice on Arrokoth detected by New Horizons is thought to be produced by successive addition of hydrogen atoms to CO ice[4,21] or radiolysis of mixed $H_2O$ and $CH_4$ ices in its early history[4,22–25]. In addition, the observed morphological structures similar to collapsed or outgassed pits at the terrain boundaries of Arrokoth[1,2] also indicate occurrence of outgassing history. These results strongly argue for the early presence of hyper volatiles in Arrokoth that could have driven the flattening process of its initial shape. Our simulations indicate that varying the driving volatile does not significantly impact the final shape (Extended Data Fig.6), although the evolution timescale varies by several times up to 1-2 orders. For instance, the timescale for CO ice driven evolution with a 5 cm dust layer composed of $R_a=10^{-6}$ m monomers is about 0.9 Myr, compared to that for $CH_4$ of about 3.1 Myr.

Other than the composition of the driving volatiles, many other factors influence the timescale of flattening, including the thickness of the insulating dust layer, the properties of dust layer particles, the size of the original object, etc. The evolution process couples these conditions and it is difficult to resolve from a modeling point of view. Overall, our simulations indicate a timescale about 1 to 100 Myr, which is orders of magnitude shorter than the dynamical age of Arrokoth (4 Gyr). This suggests that Arrokoth reached its current shape very early in its history, probably shortly after the nebula gas in the protoplanetary disk was cleared[4]. In the meantime, the timescale of tidal interaction between the two co-orbiting bodies leading to the final merger[3] is about 1-10 Myr. Therefore, although we assumed in our models that the flattening process occurred after the merging process, these two processes could have happened contemporaneously. It is possible that, if the driving sublimating ice was highly volatile on Arrokoth precursor, the two lobes have reached their respective final flattened shape before the final merger. In this case, only the neck region would have been modified by the merger process[3,4] and affected by shadowing and self-heating after the lobes combined[4]. Self-heating would dominate over shadowing in this case[4] and cause the neck region to undergo an enhanced outgassing activity after the merger, potentially resulting in its relatively high albedo and different color from the global average[4]. In any case, the relative timing of shape evolution and merging process does not affect the final shape of our model.

The New Horizons observations show a rather smooth surface of Arrokoth without the rough morphology structures commonly associated with cometary activity[2]. However, recent results from the Rosetta mission demonstrate that nominal outgassing processes do not cause the rough topography we are used to see on many comets[19,20]. On the contrary, it tends to erode large scale morphologies and smooth cometary surfaces. Large scale morphology on comets are now thought to be produced in high energy events such as outbursts , which are likely triggered by sudden increase of solar heating due to rapid decrease of heliocentric distance during dynamical evolution[7,19,20]. Therefore, violent or explosive events due to sublimation of these super volatile ices are not expected at the heliocentric distances of Arrokoth. On the other hand, when KBOs enter the Centaur phase and further become JFCs, a cycle of short term outbursts followed by a longer period of relatively stable mass loss erosion could be (re)activated[7,19,20], and successively less volatile ices could dominate the activity one by one. Our model suggests that the shape and surface structures of JFCs may not remain pristine but could have been substantially modified from their KBO stage.

Our results support the conclusion that sublimation driven mass loss is a natural mechanism for the formation of Arrokoth's flattened lobes. This process most likely occurred early in the evolution history of the body, during the presence of super volatile ices in the near subsurface layers. This could be the dominant process shaping the structure of KBOs if there were no catastrophic collision reshaping the body in their later history. Furthermore, while cold classical KBOs reserve their shape sculptured by early outgassing, the structure of Centaurs and JFCs would be further modified by the same scenario once they enter their current orbit configuration from the Kuiper Belt, under sublimation of different volatile species. We suggest that this mechanism should be taken into account in models studying planetesimal formation and the shape evolution of KBOs as well as other small icy bodies in regions where super volatile ices are expected to be present.

**Reference:**

1. Stern, S. A. *et al.* Initial results from the New Horizons exploration of 2014 MU69, a small Kuiper Belt object. *Science* **364**, (2019).
2. Spencer, J. R. *et al.* The geology and geophysics of Kuiper Belt object (486958) Arrokoth. *Science* **367**, (2020).
3. McKinnon, W. B. *et al.* The solar nebula origin of (486958) Arrokoth, a primordial contact binary in the Kuiper Belt. *Science* **367**, (2020).
4. Grundy, W. M. *et al.* Color, composition, and thermal environment of Kuiper Belt object (486958) Arrokoth. *Science* **367**, (2020).
5. Fernández, J. A. On the existence of a comet belt beyond Neptune. *Mon. Not. R. Astron. Soc.* **192**, 481–491 (1980).
6. Duncan, M., Quinn, T. & Tremaine, S. The Origin of Short-Period Comets. *Astrophys. J.* **328**, L69 (1988).
7. Zhao, Y., Rezac, L., Skorov, Y. & Li, J.-Y. The phenomenon of shape evolution from solar-driven outgassing for analogues of small Kuiper belt objects. *Mon. Not. R. Astron. Soc.* **492**, 5152–5166 (2020).
8. Porter, S. B. *et al.* High-precision Orbit Fitting and Uncertainty Analysis of (486958) 2014 MU69. *Astron. J.* **156**, 20 (2018).
9. Skorov, Y. V., Rezac, L., Hartogh, P. & Keller, H. U. Is near-surface ice the driver of dust activity on 67P/Churyumov-Gerasimenko. *A&A* **600**, A142 (2017).
10. Greenstreet, S., Gladman, B., McKinnon, W. B., Kavelaars, J. J. & Singer, K. N. Crater Density Predictions for New Horizons Flyby Target 2014 MU69. *Astrophys. J.* **872**, L5 (2019).
11. Lunine, J. I. & Stevenson, D. J. Physical state of volatiles on the surface of Triton. *Nature* **317**, 238–240 (1985).
12. Ingersoll, A. P. Dynamics of Triton's atmosphere. *Nature* **344**, 315–317 (1990).
13. Lellouch, E., De Bergh, C., Sicardy, B., Ferron, S. & Käufl, H. U. Detection of CO in Triton's atmosphere and the nature of surface-atmosphere interactions. *Astron. Astrophys.* **512**, 2–7 (2010).
14. Stern, S. A. *et al.* The pluto system: Initial results from its exploration by New Horizons. *Science* **350**, (2015).
15. Bertrand, T. & Forget, F. Observed glacier and volatile distribution on Pluto from atmosphere-topography processes. *Nature* **540**, 86–89 (2016).
16. Brown, M. E. The Compositions of Kuiper Belt Objects. *Annu. Rev. Earth Planet. Sci.* **40**, 467–494 (2012).
17. Wierzchos, K., Womack, M. & Sarid, G. Carbon Monoxide in the Distantly Active Centaur (60558) 174P/Echeclus at 6 au. *Astron. J.* **153**, 230 (2017).
18. Womack, M., Sarid, G. & Wierzchos, K. CO and other volatiles in distantly active comets. *Publ. Astron. Soc. Pac.* **129**, 1–19 (2017).
19. Vincent, J. *et al.* Constraints on cometary surface evolution derived from a statistical analysis of 67P's topography. *Mon. Not. R. Astron. Soc.* **469**, S329-S338 (2017).
20. Ramy El-Maarry, M. *et al.* Surface changes on comet 67P/Churyumov-Gerasimenko suggest a more active past. *Science* **355**, 1392–1395 (2017).
21. Bosman, A. D., Walsh, C. & Van Dishoeck, E. F. CO destruction in protoplanetary disk midplanes: Inside versus outside the CO snow surface. *Astron. Astrophys.* **618**, 1–19 (2018).

22. Hodyss, R., Johnson, P. V., Stern, J. V., Goguen, J. D. & Kanik, I. Photochemistry of methane-water ices. *Icarus* **200**, 338–342 (2009).
23. Moore, M. H. & Hudson, R. L. Infrared Study of Ion-Irradiated Water-Ice Mixtures with Hydrocarbons Relevant to Comets. *Icarus* **135**, 518–527 (1998).
24. Pearce, M. P. *et al.* Formation of methanol from methane and water in an electrical discharge. *Phys Chem Chem Phys* **14**, 3444–3449 (2012).
25. Wada, A., Mochizuki, N. & Hiraoka, K. Methanol Formation from Electron-irradiated Mixed $H_2O/CH_4$ Ice at 10 K. *Astrophys. J.* **644**, 300–306 (2006).
26. Sullivan, C. B. & Kaszynski, A. {PyVista}: 3D plotting and mesh analysis through a streamlined interface for the Visualization Toolkit ({VTK}). *J. Open Source Softw.* **4**, 1450 (2019).
27. Takahashi, Y., Busch, M. W. & Scheeres, D. J. Spin state and moment of inertia characterization of 4179 Toutatis. *Astron. J.* **146**, 95 (2013).
28. Zhao, Y. *et al.* Orientation and rotational parameters of asteroid 4179 Toutatis: new insights from Chang′e-2's close flyby. *Mon. Not. R. Astron. Soc.* **450**, 3620–3632 (2015).
29. Keller, H. U., Mottola, S., Skorov, Y. & Jorda, L. The changing rotation period of comet 67P/Churyumov-Gerasimenko controlled by its activity. *Astron. Astrophys.* **579**, L5 (2015).
30. Zhao, Y., Hu, S., Wang, S. & Ji, J. Using a volume discretization method to compute the surface gravity of irregular small bodies. *Chin. Astron. Astrophys.* **40**, 45–53 (2016).
31. Samarasinha, N. H. & Belton, M. J. Long-term evolution of rotational states and nongravitational effects for Halley-like cometary nuclei. *Icarus* **116**, 340–358 (1995).
32. Samarasinha, N. H. Preferred Orientations for the Rotational Angular Momentum Vectors of Periodic Comets. in *BAAS* vol. **29,** 743 (1997).
33. Choi, Y.-J., Cohen, M., Merk, R. & Prialnik, D. Long-Term Evolution of Objects in the Kuiper Belt Zone – Effects of Insolation and Radiogenic Heating. *Icarus* **160**, 300–312 (2002).
34. Prialnik, D., Sarid, G., Rosenberg, E. D. & Merk, R. Thermal and Chemical Evolution of Comet Nuclei and Kuiper Belt Objects. *Space Sci. Rev.* **138**, 147–164 (2008).
35. Skorov, Y. & Blum, J. Dust release and tensile strength of the non-volatile layer of cometary nuclei. *Icarus* **221**, 1–11 (2012).
36. Gundlach, B. & Blum, J. A new method to determine the grain size of planetary regolith. *Icarus* **223**, 479–492 (2013).
37. Marohnic, J. C. et al. Constraining the final merger of contact binary (486958) Arrokoth with soft-sphere discrete element simulations. Icarus 113824 (2020).

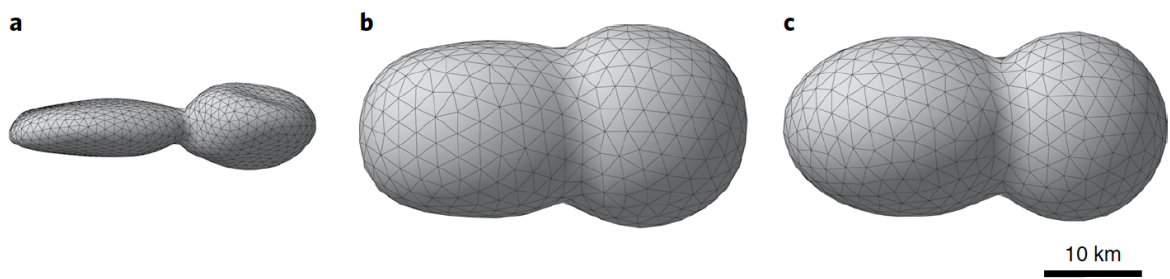

**Figure 1. Shape models of the Arrokoth KBO as observed today and the analogues of a parent body derived from our modeling.** Fig. 1a shows the shape of Arrokoth as derived from New Horizons' observations[1], referred to as SHAPE-ORIG in our work. Starting with this digital

terrain model, we perform inverse simulation (Methods) to obtain the SHAPE-REV shown in Fig. 1b. The backward erosion process is simply implemented as negative illumination driven sublimation, such that the body gains mass during its orbital and spin motion. The self-consistency of volume loss/gain in the forward and reverse direction due to sublimation mass loss in our model has been described elsewhere[7]. In the reverse erosion mode, the simulations are performed with pure $CH_4$ ice (zero thermal inertia) and were run for about 1 Myr to obtain the body SHAPE-REV shown here. Fig. 1c shows a mathematically idealized body that approximates SHAPE-REV, keeping the volume and axis ratios of both lobes. It is formed by joining a sphere and an oblate shape, and it is taken as the parent body of Arrokoth in our simulations. This figure is shown from the view along the +y axis, and it represents the shapes in their relative sizes in the same scale.

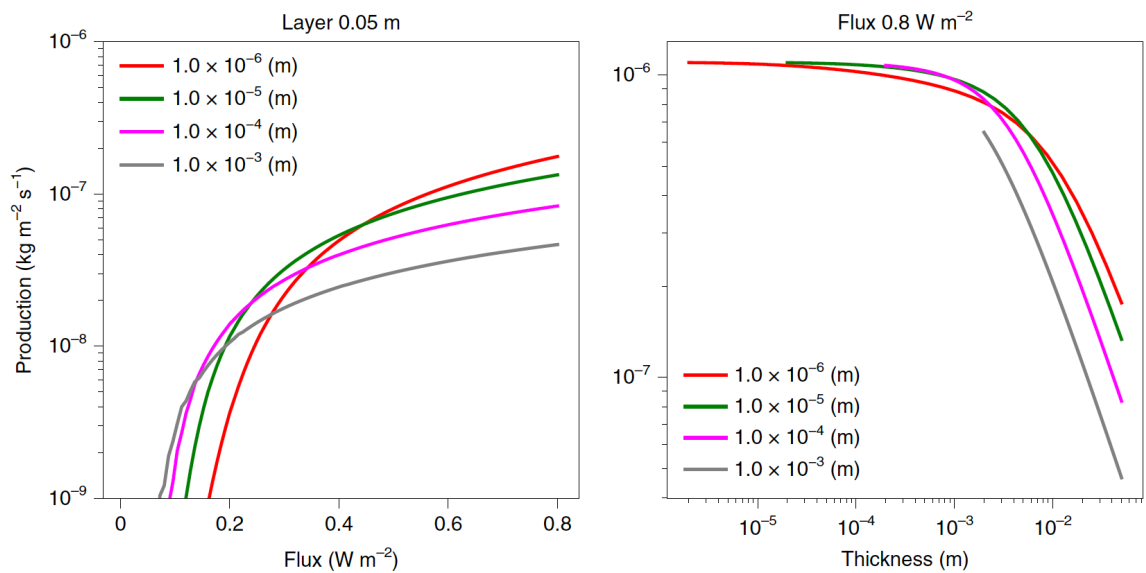

**Figure 2. $CH_4$ mass loss rate of different dust layer models.** This figure illustrates outgassing rate variability with solar flux for a dust layer of 5 cm thickness (left panel). Each curve in the plot corresponds to a dust layer built from different sized monomers as labeled. The right panel shows outgassing rate versus layer thickness for conditions of a near maximum solar flux at Arrokoth's perihelion. This plot illustrates two important points. First, the strongest sublimation occurs for cases of thin layers, especially made of small grains, as expected. The thicker the layer, the less heat can reach the subsurface ice which leads to a rapid reduction in mass loss rate. However, for cases of thick layers made from smaller monomers, which provide more contacts per area, more heat is conducted to ice which leads to a stronger sublimation rate than for layers made of larger monomers. This is true as long as the solid conductivity component dominates the total conductivity coefficient which can be argued to be a general case for regolith at these heliocentric distances. The other important dust layer characteristics to consider here is the permeability which varies ~ $R_α/L$ according to the Knudsen formula, where L stands for a layer thickness and $R_α$ for a monomer size. At the right panel of the figure it is worthwhile to point out how different curves of mass loss intersect at different locations in the parameter space ($R_a$, L); It indicates that identical objects of the same orbital parameters and/or spin axis may experience slightly different mass loss distributions depending on different dust grain sizes. This is an important point which generalizes our results and highlights the potential importance of the mass loss process on the overall morphology.

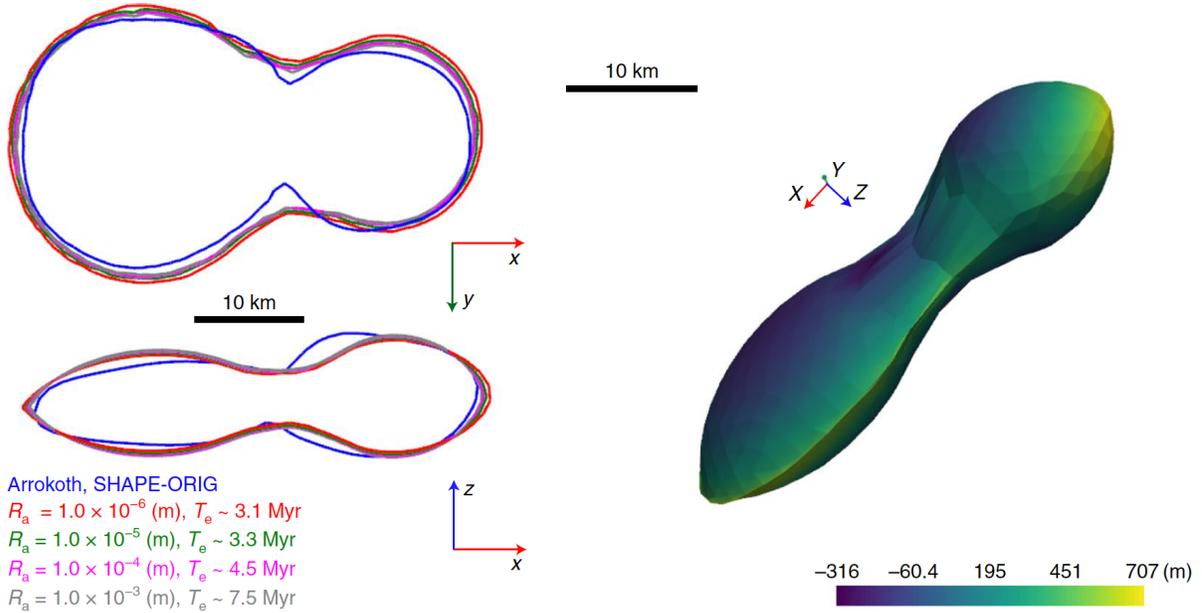

**Figure 3. Final shapes of Arrokoth analogues simulating mass loss evolution with different dust layer properties.** The left panel shows cross-sections of evolved bodies that reached approximately the same volume as SHAPE-ORIG shown in blue outline for reference, the timescales $T_e$ are shown in the key. The colored lines [red, green, magenta, gray] identify the different cases of monomer sizes, $R_a$, as labeled, that make up the model dust layer of 5 cm thick. A bulk porosity of 50% and a volume dust-to-ice ratio of two are assumed in the model. The evolved bodies all reproduce the overall structure and shape characteristics of the SHAPE-ORIG irrespective of the specific dust layer properties. Nevertheless, there are differences in some details which result from different dust layer characteristics. The right panel shows another visualization of the differences in the morphology due to dust layer effect on the mass loss rate. The color scale of the shape indicates radial differences of facet centers between final shapes with a dust layer built from monomer size $R_a=10^{-3}$ and $R_a=10^{-6}$, corresponding to the gray and red cross-section in the left panel. We observe that volume lost rapidly from regions experiencing direct solar flux for $R_a=10^{-6}$, however, the poorly illuminated equatorial regions suffer a weaker volume loss in the evolutionary period before the body reaches the volume of SHAPE-ORIG. The results demonstrate

that flattening is a natural outcome because of Arrokoth's large obliquity (99°) and small eccentricity, and is not strongly dependent on the dust layer properties, e.g. thickness or monomer size of dust particles. Furthermore, the "dust mantle" could also be a mixture of dust particles and non-volatile ices, such as water ice. This will not change the conclusion of the mass loss driven flattening[34,35], but only change the evolution timescale for the body to reach SHAPE-ORIG.

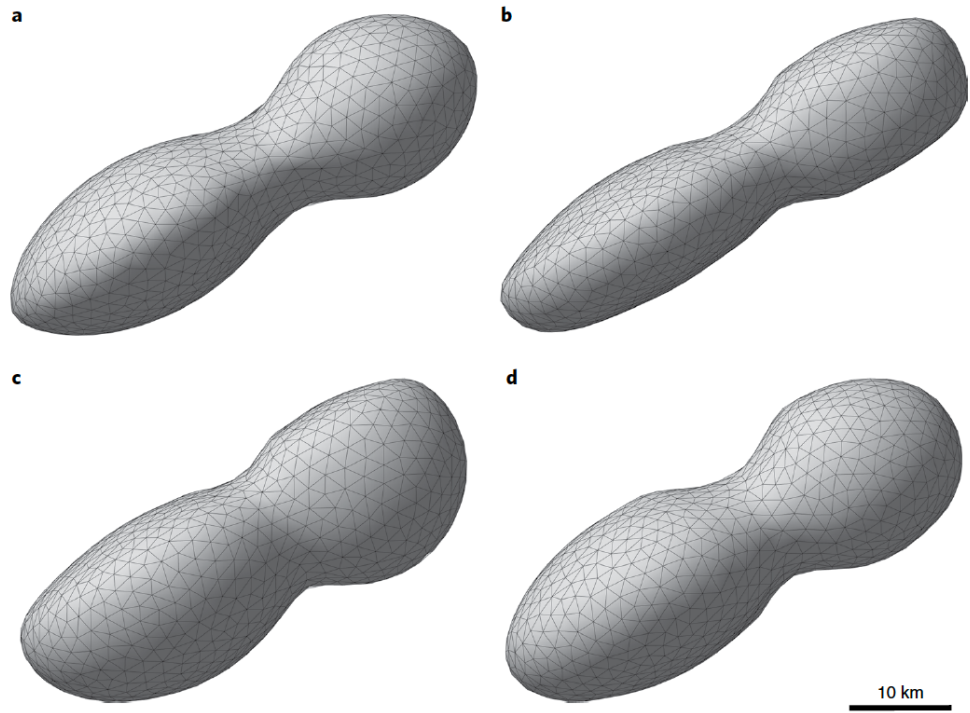

**Figure 4. Shape evolution results taking into account gravity, and tensile stress effects on the mass loss distribution: a) nominal case; b) tensile stress threshold of 0.017 Pa; c) tensile stress threshold of 0.02 Pa; d) accounting for gravity.** In panel a we show the final shape under nominal conditions of mass loss by $CH_4$ sublimation for a dust layer of 5 cm thickness made of $R_a=10^{-5}$ m monomers. Panels b and c show resulting shapes due to mass loss occurring only when subsurface pressure exceeds a threshold of 0.017 Pa and 0.02 Pa respectively. The flattening of the two lobes in panel b is more pronounced compared to the nominal case because the equatorial region experiences a weaker mass loss. In addition, the north/south differences in flattening are amplified as a result, and that is despite the small eccentricity of the orbit of Arrokoth. In contrast, the final shape shown in panel c is flattened only on the summer hemisphere as a result of increasing the pressure threshold by only 0.003 Pa. The eccentricity and the threshold combine into an effect where the winter hemisphere experiences significantly weaker mass loss. This particular simulation did not reach the approximate volume of Arrokoth during the evolution time of 5 Myr. In panel d we show a case where mass loss occurs only when the pressure gradient force is larger than acceleration due to gravity (including centrifugal acceleration). This condition

creates yet a different final shape compared to the other ones. Although flattening of lobes still occurs, an additional morphological feature resembling a pinnacle or hill develops at areas of gravitation highs. In addition, its size is not symmetrical between north and south hemispheres. Although we selected these thresholds subjectively for the particular dust layer in order to illustrate these key points, it is certain that a combination of the thermophysical, mechanical and ice composition of the real Arrokoth could be crucial for shaping the morphology. It is remarkable that even for initially round objects on nearly a circular orbit the sublimation driven mass loss may result in a non-round shape exhibiting strong north/south asymmetry depending on the dust (and ice) properties and their distribution.

**Methods**

**Shape models of Arrokoth and shape evolution modeling**
The shape of Arrokoth was initially derived from New Horizons' flyby images[1] as bilobed with respective dimensions of 20×20×7 km and 14×14×10 km (SHAPE-ORIG, Fig. 1a), which is used as initial shape model in our simulations. Later it was slightly updated[2] to 20.6×19.9×9.4 km and 15.4×13.8×9.8 km, with uncertainties of 0.5 km × 0.5 km ×2.0 km.. The new bi-lobed shape model shows a slightly less flattened structure, larger volume, and the neck region seems to be smoother. Nevertheless, the pronounced lenticular shape of both lobes is maintained, and in our simulations such volume would be reached perhaps in a shorter timescale. Therefore, the update of Arrokoth's shape does not affect our conclusions.

The shape evolution model (MONET) is detailed in our previous work[7], including description of the numerical procedure translating mass loss to a decrease in relevant facet's depth. Here we present only the general idea of implementation of the depth change of facets resulting from material removal. The displacement of the *l*-th vertex $\boldsymbol{D}_l^v$ is a combination of depth changes $D_k^f$ of all *m* number of facets sharing it, weighted by the depth changes itself[7],

$$\boldsymbol{D}_l^v = -\frac{\sum_{k=1}^{m}(D_k^f)^2 \boldsymbol{n}_k}{\sum_{k=1}^{m} D_k^f},$$

where $\boldsymbol{n}_k$ is the normal direction of the *k*-th facet. The reverse erosion procedure described in this work from which we obtain the SHAPE-REV from SHAPE-ORIG is implemented simply by changing the negative sign of this expression into positive. The backward propagation of the evolution model is validated in our previous work[7].

However, we want to note that the shape modification procedure (i.e. reduction of radius vector of a facet upon accumulated mass loss) for an arbitrary 3D object is still a topic of research. In our scheme, a particular challenge is shrinking volume with concavities (e.g. the neck region of Arrokoth), which leads to intersecting facets (triangles) after some time. One way to handle this is by re-meshing the triangular mesh, where we experimented extensively with relevant parameters with the help of PyVista and pyacvd packages[26] to accomplish this. Nevertheless, we found that even these state-of-the-art re-meshing techniques cannot exactly preserve volume, and/or result in excessive smoothing of sharp features. Therefore, the remeshing procedure is not used in automatic mode, but applied only once at an empirically determined step usually toward the end of simulation in order to reduce triangular facet distortions resulting from the volume changes.

**Spin State Modeling**

We investigate the role of sublimation torques self-consistently under conditions of the evolving nucleus shape derived from our model. Based on the orbital parameters[8], Arrokoth's obliquity and longitude in its orbital frame are 99.2° and -20.5 °, respectively. The large obliquity and small longitude indicate that the spin axis of this body lies in the direction close to the perihelion of its orbit. Suppose matrix R is employed to describe the orientation and spin state of the nucleus in space in terms of 3-1-3 Euler angles ($\alpha$, $\beta$, $\gamma$). The transformation of a vector in the inertial coordinate system, $r$, to that in the nucleus body-fixed system, $R$, reads:

$$\boldsymbol{R} = \mathrm{R}\boldsymbol{r} = \mathrm{Rz}(\gamma)\mathrm{Rx}(\beta)\mathrm{Rz}(\alpha)\boldsymbol{r},$$

where Rx and Rz are the standard rotation matrices for right-handed rotations around the x and z axes. In this work, we perform all the simulations in an ecliptic frame, the corresponding values of the first two Euler angles ($\alpha$, $\beta$) are 42.5° and 98.1° derived from its right ascension and declination[2]. These values are used as initial states in our simulations. Because the rotational period is small compared to the orbital period or the shape change timescale, the initial value of rotation angle $\gamma$ is unimportant and is chosen to be 0.

To investigate how spin states evolve, we start with principal spin state under the assumption of a homogeneous body and solve the Euler equations[27,28] taking into account only the effect induced by sublimation. The current spin state of Arrokoth is used as the initial parameters, and the external torque $\boldsymbol{L}$ has a form of:

$$\boldsymbol{L} = -\sum_i \frac{\mathrm{d}m_i}{\mathrm{d}t}\boldsymbol{r}_i \times \boldsymbol{v}_i,$$

where $m_i$ is the mass loss of the $i$-th facet, $\boldsymbol{r_i}$ is the vector from nucleus' mass center to facet's center, and $\boldsymbol{v_i}$ is gas ejection velocity which is considered to be in the direction of facet's normal and computed as[29]

$$\boldsymbol{v_i} = \eta \sqrt{\frac{8RT_g}{m\pi}},$$

where $m$, $R$ and $T_g$ are molar mass of volatile molecules, gas constant and gas temperature, respectively. The constant η which represents the efficiency of impulse transfer has a value of 0.77[29]. In Extended Data Fig. 5 we show torque efficiency[29] distribution for both SHAPE-SYN and SHAPE-REV.

We use the Runge-Kutta-Fehlberg (RKF78) method to solve the Euler equation. A finite element method (FEM)[30] is employed to determine the values of inertia tensor for the initial shape, and it is then re-calculated after each shape modification step. Because of the low eccentricity and large obliquity of Arrokoth, the north-south asymmetry often developed due to sublimation driven mass loss is insignificant. For this, and the computational cost reasons we ignore the change of products of inertia Ixz, Iyz in the inertia matrix in our simulations. Since the SHAPE-SYN is symmetric with respect to the x-y plane, Ixz, Iyz remain zero during the entire evolution. However, in the case of SHAPE-REV, the symmetry conditions are not met and the off-diagonal of the inertia tensor is expected to be non-zero. We obtain the initial values of components for SHAPE-REV, (Ixx, Iyy, Izz , Ixy , Ixz ,Iyz) = ($3.9525 \times 10^{23}$, $9.7404 \times 10^{23}$, $1.0419 \times 10^{24}$, $3.1737 \times 10^{21}$, $1.8406 \times 10^{22}$; $1.9410 \times 10^{21}$ kgm$^2$) with the scale of elements about 500 m. This yields a deviation of angular momentum direction from the spin-axis (z-axis) by about 1º. This offset could be a result of the heterogeneity of density distribution, but such a small value is not expected to be observable by a flyby mission. In addition, we believe this value is also contaminated by the errors introduced by the FEM calculation, therefore, we choose to ignore these terms for SHAPE-REV when evaluating the spin and orientation evolution.

It is a challenge for numerical integration schemes to evaluate sublimation torques for objects at such large heliocentric distances since the spin period is far less than the orbital period: $P_{rot} \ll P_{orb}$. Due to the requirement of self-consistent mass loss with 3D shape modification, no reasonable analytical solution can be applied so far. On the other hand, the integration of the numerical process is so computationally intensive that even on multi-core clusters that we cannot cover extended time periods (~1-10 Myr). However, we present a tendency of the simulated spin state evolution for around 0.05 to 0.1 Myr and provide a supporting argument for validity to extrapolate these to longer timescales.

From our simulation results (Extended Data Fig. 3, 4, 5), and from the fact that Arrokoth's spin axis lies close to its perihelion direction, we may conclude that changes in the orientation of spin axis and spin rate would not be very significant even if we extrapolate these trends to the entire timescale of 1 Myr. Although the moments of inertia are getting smaller due to mass loss which should tend to accelerate the spin state change, as the spin axis is getting close to orbital plane, the changing rate decreases as the net torque in the inertial frame is minimized[31,32] (Extended Data Fig. 5). These competing mechanisms will most likely keep the change within a small range. Compared to the timescale of the shape modification process, the simulation results of spin state evolution show an insignificant effect induced by sublimation torques. This is true for both the orientation and the spin period change during the mass loss process.

The current SHAPE-ORIG constrains the evolution time, therefore also the sublimation torques perturbation to the spin and orientation of Arrokoth in our simulations. Although these results are done under assumption of pure $CH_4$ ice sublimation, which provides the upper limit on the sublimation rate. For different volatile species or dust layer properties the evolution time may be shorter/longer while the associated sublimation torques would in turn act over a shorter/longer time period respectively. Therefore, simulations always show a similar level of perturbations as discussed, independent of evolution time (to the first order).

In the spin state simulations, we are not able to perform parametric study of different volatile gasses, dust-layers, or density distributions., however, the resulting spin state change would be of similar magnitude to our simulation case. Furthermore, a long standing issue when simulating torques on asteroids, comets, and KBOs is the assumption of rigid body dynamics. Non-rigidity of internal structure would change the timescale of NPA excitation for small bodies, but the necessary data to relax this assumption are also needed to bring models closer to reality.

In summary, the YORP effect plays a negligible role in altering Arrokoth's spin state[1,3], collision perturbations are also rare as indicated from its lightly cratered surface[1,2,10], therefore, we conclude that the spin state of Arrokoth didn't change significantly during its evolutionary history even taking sublimation torques into consideration (under the assumption of homogeneous body). Hence, with a stable obliquity of around 99°, the long-term mass loss could be simulated by not taking into account spin-orientation changes during the evolution. This significantly simplifies the computations and increases efficiency.

**Modeling Assumptions**

There are two highly idealizing assumptions in our model, a) the homogeneous distribution of ice/dust properties in the entire volume of the body, and b) constant layer thickness during the mass loss evolution. These are the most difficult to relax due to our general lack of the ab-initio knowledge of dust lifting processes, transport and subsequent weathering, and unknown interior and microphysical dust/ice properties of this body. Therefore, transient thermal modeling does not seem to be justified, especially for large scale morphology evolution due to the lack of knowledge of volumetric distribution of microphysical dust and ice properties. It is only an educated guess to attempt to assess the impact of these coupled issues on the evolution times of our simulations. We can get some idea from the tensile strength simulations shown Fig. 4.

The assumption of a dust layer with constant thickness during the mass loss duration implies that our model does not quench evolution from first principles, but it is stopped when it reaches the (approximate) volume of SHAPE-ORIG. An idealized slowly thickening dust layer could gradually decrease the mass loss rate and finally stop it. It is also possible that internal heating[33] could have provided an internal heat source to drive sublimation and subsequent recondensation of super volatiles closer to the surface where the solar illumination could drive mass loss flattening and continue to do until this "layer" of volatile ice was depleted, hence stopping the mass loss. However, we cannot provide a definite answer to the question that how the sublimation stopped when the shape reached SHAPE-SYN, because of the lack of knowledge about the formation and evolution of these bodies from both macroscopic and microcosmic point of view. In particular, the physics of dust lifting or settling, dust layer growth and heat transport through it is an extremely complex problem. Further theoretical and numerical investigation relying on future data about physics of ices and dust is required. At this time, it is out of the scope of this work.

In addition to the illumination driven sublimation shaping these bodies, the future modeling should be considered together with the chemical and thermal evolution of the interiors as well[33,34]. The processes of internal heating (due to radiogenic decay with subsequent amorphous ice crystallization) can also significantly influence the internal structure and composition of the icy body[35,36], and could even lead to shorter evolution time for flattening.

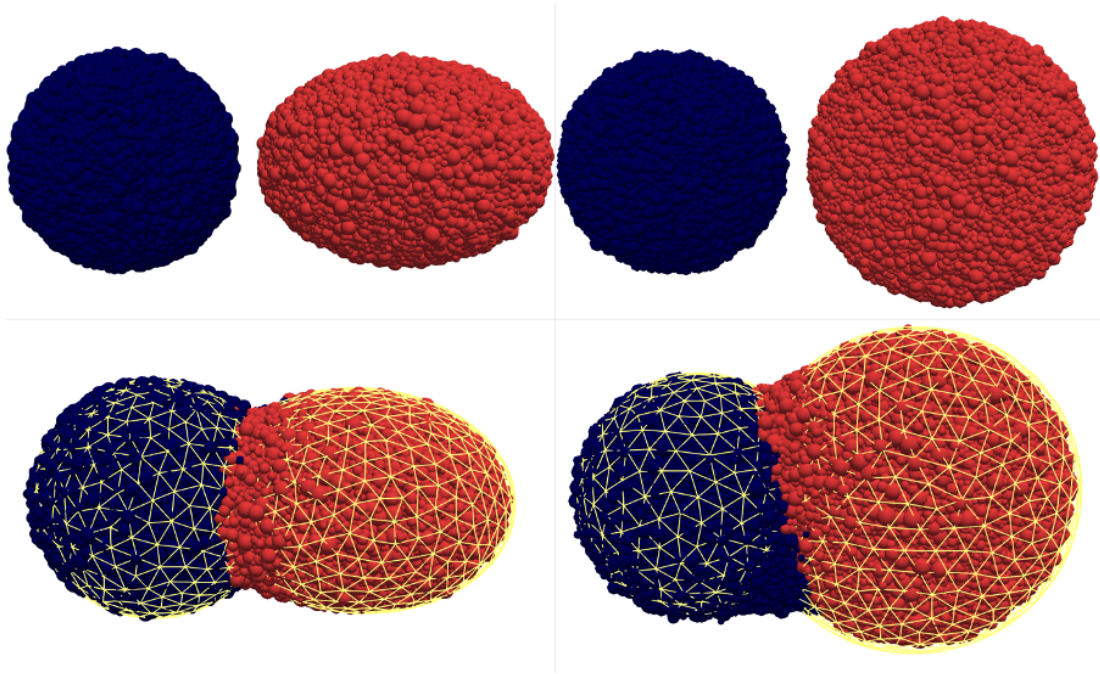

**Extended Data Figure 1. Shape configuration produced by a low-speed collision of a spherical and an oblate body.** The soft-sphere discrete element method implemented by N-body code pkdgrav is used to simulate the merging process of two planetesimals that formed the precursor body of Arrokoth[3,25]. The primary and secondary bodies in the upper panels (oblate and spherical shape, respectively) are of the same scale as the lobes of SHAPE-SYN. The two lobes are constructed with 40,000 and 25,775 spherical particles with a power-law distribution of radii ranging from 309 m to 927 m using a -3 index, and an inter-particle cohesion of 27500 Pa is assumed[1]. A set of material parameters are selected to obtain a friction angle of 40° and the bulk density[3] is set to be 0.47 g/cm$^{-3}$. The resulting contact binary shown in the lower panels is obtained with impact speed v = 5 m/s and impact angle α= 45° (measured with respect to the vertical at the impact point), which is close to the SHAPE-SYN. The upper panels show the two original objects from -y axis and +z axis viewpoints, in which blue and red particles belong to the spherical and oblate body, respectively. The bottom panel shows the resulting contact binary for the same viewing geometry, and the wireframe in yellow outlines the SHAPE-SYN we build from the inverse evolution result. This simulation result shows a good agreement and indicates that our idealized bilobed body is consistent with physics of slow collisions, and, as already noted, it is taken as our starting shape in our evolution model.

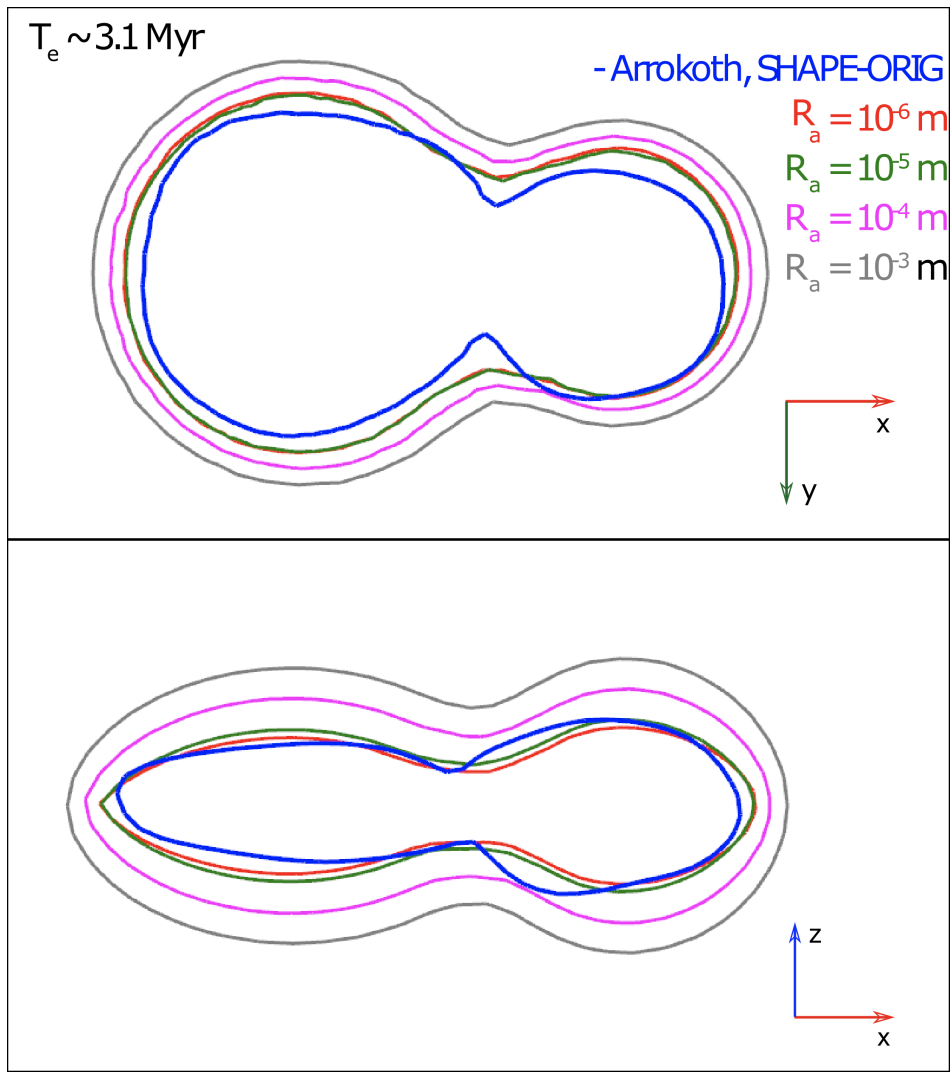

**Extended Data Figure 2. Results illustrating the time evolution of Arrokoth analogues for different dust layer properties at a fixed point in time.** For easier visual comparison we plot only extracted cross sections aligned at the body axes. The blue outline belongs to Arrokoth (SHAPE-ORIG), shown for a reference. The other colored outlines [red, green, magenta, gray] are for cases with different dust layers, each 5 cm in thickness but varying monomer particle sizes, $R_a$, from $10^{-6}$ to $10^{-3}$ meters (as labeled in the figure). The curves show results extracted at the same time instant of about 3.1 Myr, which is when the scenario $R_a = 10^{-6}$ m (red curve) reaches the approximate volume of Arrokoth. On the other hand, the layer constructed from the largest monomers (grey curve) conducts little heat to the ice and provides the weakest mass loss. Under these conditions, it would take another 4.4~Myr to reach the same volume as SHAPE-ORIG.

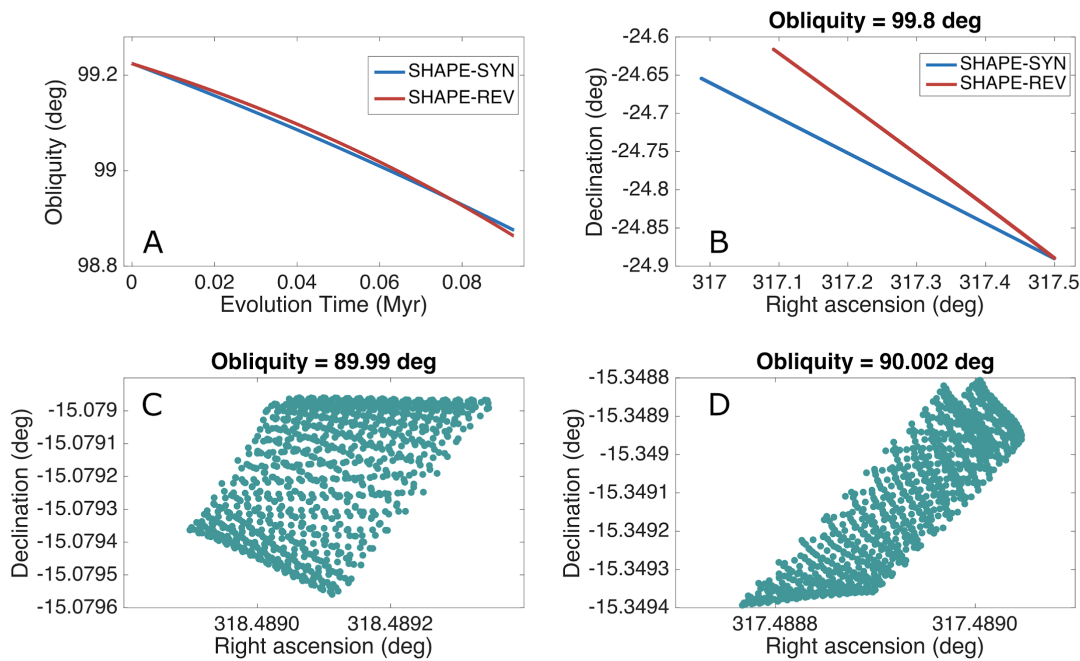

**Extended Data Figure 3. Simulations of obliquity and orientation evolution taking into account sublimation torques.** These simulations are performed with reduced assumption of pure $CH_4$ icy bodies in order to get the upper limit of the sublimation torques. This reduces the time to reach the volume of the SHAPE-ORIG from the initial states (SHAPE-SYN and SHAPE-REV) to about 1 Myr. Due to the computationally intensive integral needed to solve the Euler equation, we present a tendency of the simulated spin state evolution for around 0.05 to 0.1 Myr. In panel A we show the obliquity change for both initial shapes as a function of time (for approx. 320 orbits or 0.1 Myr), starting with the current obliquity of Arrokoth. The relevant movement of the spin axis' orientation in the space is shown in panel B in terms of right ascension and declination (Ra, Dec). The initial (Ra, Dec) state is in the lower right corner of the plot. These simulations show that obliquity, Ra. and Dec. changes are within 0.4, 0.3 and 0.5 deg respectively. The spin axis has a tendency to stay near the orbital plane, which is demonstrated in panels C and D, for two cases of obliquity close to 90° (89.99° and 90.002°) starting with SHAPE-SYN. The orientation changes of the spin axis are presented again in Ra. Dec. space for a simulation time of 200 orbits (about 0.06 Myr), during which the variation is less than $10^{-3}$ deg for both angles. A strong stability of orientation is preserved when the spin axis lies closely to the orbital plane and perihelion because the net torque in the inertial frame is minimized[30,31].

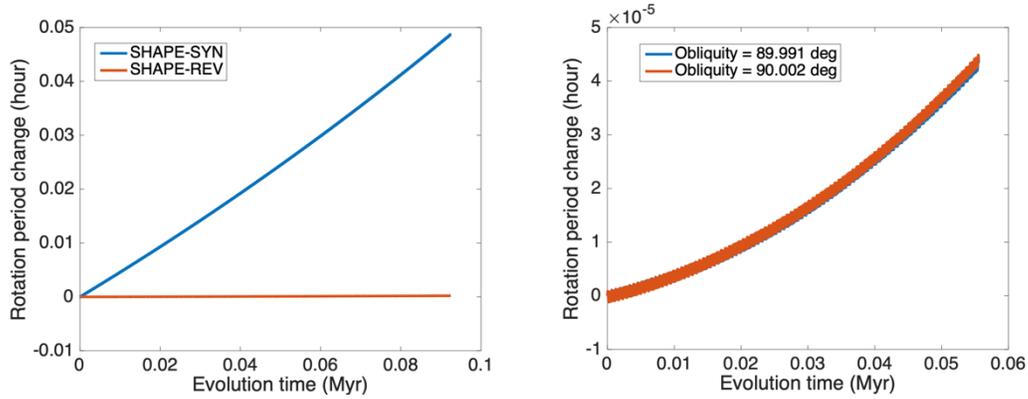

**Extended Data Figure 4. Simulation results estimating the magnitude of spin period change.** (Left) Time evolution of rotational period change for SHAPE-SYN and SHAPE-REV assuming current Arrokoth obliquity. The SHAPE-REV body experiences stronger changes in spin period by about two orders of magnitude during simulation because it's more asymmetrical compared to the SHAPE-SYN (see also Supplementary Fig. 5). (Right) simulations done for SHAPE-REV for two cases of obliquity close to 90º in which case the spin period does change significantly slower, at change of ~$10^{-5}$ hrs after 0.6 Myr. Here, combined with the result shown in Supplementary Fig. 5, we can conclude that if the external gravitational torque from planet encounters could be ignored[1,3], then the spin state of Arrokoth didn't change significantly during its evolution history even when taking sublimation torques into consideration (under the assumptions of homogeneity).

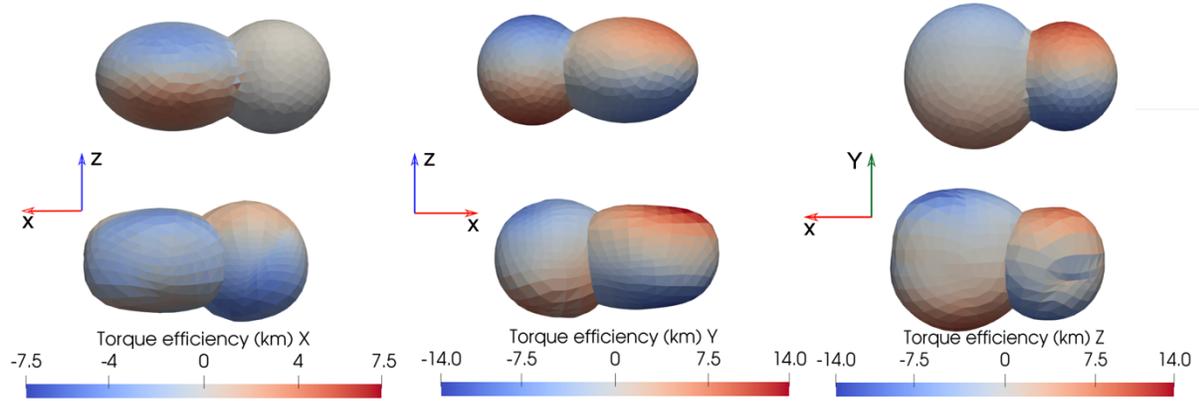

**Extended Data Figure 5. Torque efficiency of SHAPE-SYN (upper panels) and SHAPE-REV (lower panels).** Here we provide additional information on surface torque efficiency $-r_i \times \hat{n}_{vi}$ in x, y and z direction for both initial shapes to assess whether to expect significant sublimation driven torques on the body. Here, $r_i$ and $\hat{n}_{vi}$ are the radius and the unit vector in the direction of gas ejection of i-th facet. The distribution of positive (red) and negative (blue) values of torque efficiency indicating increasing or decreasing of the spin rate along the corresponding spin axis, are perfectly "symmetric" on the surface of SHAPE-SYN. When the body is in a circular orbit with an obliquity of 90º (or 0º), the orbital averaged solar flux distribution on its surface is even between "red" and "blue" regions, resulting in a balance between the torques induced by mass loss in both regions This balance could not be perfectly achieved for SHAPE-REV since it exhibits a small degree of asymmetry, but the resulted non-zero torque is also of small absolute value in the case of circular orbit and an obliquity of 90º. Since Arrokoth has an obliquity close to 99º and small orbital eccentricity, the illumination distribution on its surface is evenly distributed in positive and negative regions and, hence, leads to a low magnitude of the sublimation induced torques. The red, green and blue arrows point in the direction of +x, +y and +z respectively, as labeled in the plots.

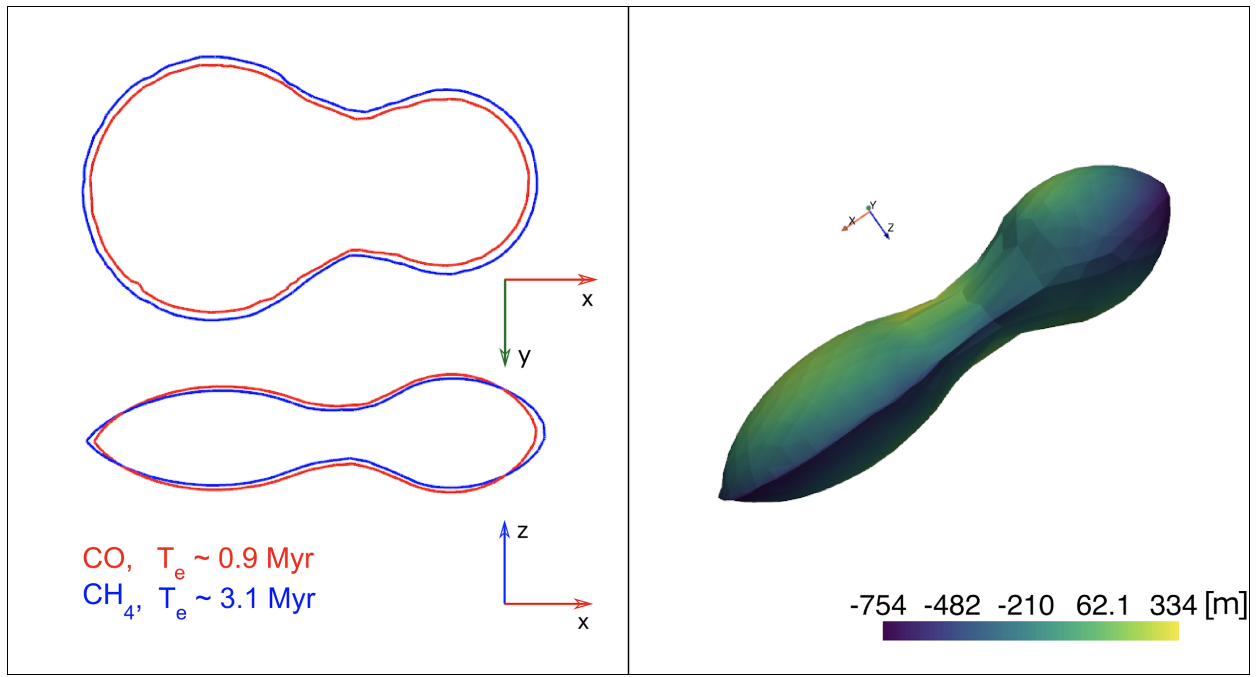

**Extended Data Figure 6. Comparison of evolution simulations for two different ices.** The red outline belongs to the CO ice sublimation case, while the blue one is for $CH_4$. In both cases the properties of the dust layer are the same with a thickness of 5 cm and $R_a = 10^{-6}$ m. It takes about 0.9 Myr to reach the volume of SHAPE-ORIG when CO is the major volatile species, while about 3.1 Myr for $CH_4$ ice sublimation. In the left panel the results are shown as cross-sections through equatorial and polar regions. In the right panel we show the final 3D body with color scale denoting radial differences between facet centers of the two cases ($CH_4$ minus CO case). From the point of view of the flattening process of the two lobes we conclude that the results are nearly identical, hence, independent of the type of ice in question. However, a faster volume loss due to higher CO volatility produces less erosion in the polar regions but higher at the equatorial areas in the time it takes to reach the SHAPE-ORIG volume. The principle of this effect follows from the fact that the CO to $CH_4$ volatility in the less illuminated equatorial region is higher than the time ratio needed to reach the final shape, which is not the case for areas under nearly constant illumination.